\documentstyle[preprint,floats,epsf,tighten,aps]{revtex} 

\def\lqcd{\Lambda_{\rm QCD}}
\def\qsh{\hat q^2}
\def\qcut{q_0}
\def\Qcut{Q_0}
\def\mbar{{\bar m}_B}

\begin{document}

\preprint{\vbox{ \hbox{UTPT-00-02} \hbox{FERMILAB-Pub-00/039-T}
  \hbox{hep-ph/0002161} \hbox{} }}

\title{A Model Independent Determination of $|V_{ub}|$}

\author{Christian W.\ Bauer,$^{a}$ 
  Zoltan Ligeti,$^{b}$ and Michael Luke$^{\,a}$}

\address{ \vbox{\vskip 0.truecm}
  $^a$Department of Physics, University of Toronto, \\
    60 St.\ George Street, Toronto, Ontario, Canada M5S 1A7 \\ [8pt]
  $^b$Theory Group, Fermilab, P.O.\ Box 500, Batavia, IL 60510 }

\maketitle

\begin{abstract}%
It is shown that measuring the lepton invariant mass spectrum in inclusive
semileptonic $\bar B\to X_u \ell\bar\nu$ decay yields a model independent
determination of $|V_{ub}|$.  Unlike the lepton energy and hadronic invariant
mass spectra, nonperturbative effects are only important in the resonance
region, and play a parametrically suppressed role when ${\rm d}\Gamma / {\rm
d}q^2$ is integrated over $q^2 > (m_B-m_D)^2$, which is required to eliminate
the charm background.  Perturbative and nonperturbative corrections are
presented to order $\alpha_s^2\beta_0$ and $\Lambda_{\rm QCD}^2/m_b^2$, and the
$\Lambda_{\rm QCD}^3/m_b^3$ corrections are used to estimate the uncertainty in
our results.  The utility of the $\bar B\to X_s \ell^+\ell^-$ decay rate above
the $\psi(2S)$ resonance is discussed.

\end{abstract}

\newpage

A precise and model independent determination of the Cabibbo-Kobayashi-Maskawa
(CKM) matrix element $V_{ub}$ is important for testing the Standard Model at
$B$ factories via the comparison of the angles and the sides of the unitarity
triangle.  The first extraction of $|V_{ub}|$ from experimental data relied on
a study of the lepton energy spectrum in inclusive charmless semileptonic $B$
decay~\cite{CLEO1}.  Recently $|V_{ub}|$ was also measured from exclusive
semileptonic $\bar B\to \rho \ell \bar\nu$ and $\bar B\to \pi \ell \bar\nu$
decay~\cite{CLEOexcl}, and from inclusive decays using the reconstruction of
the invariant mass of the hadronic final state~\cite{Vubmass}.  

These determinations suffer from large model dependence.  The exclusive
$|V_{ub}|$ measurements rely on form factor models or quenched lattice
calculations at the present time.\footnote{A model independent determination of
$|V_{ub}|$ from exclusive decays is possible without first order heavy quark
symmetry or chiral symmetry breaking corrections~\cite{lw}, but it requires
data on $\bar B\to K^* \ell^+\ell^-$.  A model independent extraction is also
possible from decays to wrong-sign charm \cite{wrongsigncharm}, but this is
very challenging experimentally.  See also \cite{KPY} for a discussion of
extracting $|V_{ub}|$ from a comparison of photon spectra in $B$ and $D$
radiative leptonic decays.} Inclusive $B$
decay rates are currently on a better theoretical footing, since they can be computed model
independently in a series in $\lqcd/m_b$ and $\alpha_s(m_b)$ using an operator product
expansion (OPE)~\cite{CGG,incl,MaWi,Blok}.  However, the predictions of the OPE are only
model independent for sufficiently inclusive observables, while the $\bar
B\rightarrow X_u\ell \bar\nu$ decay rate can only be measured by imposing
severe cuts on the phase space to eliminate the $\sim 100$ times larger $\bar
B\rightarrow X_c\ell\bar\nu$ background. For both the charged lepton and
hadronic invariant mass  spectra, these cuts spoil the convergence of the OPE,
and the most singular terms must be resummed into a nonperturbative $b$ quark
distribution function\cite{shape}.  While it may be possible to extract this
from the photon spectrum in $B\to X_s\gamma$ \cite{shape,LLR}, it would clearly
be simpler to find an observable for which the OPE did not break down in the
region of phase space free from charm background.  In this Letter we show that
this is the situation for the lepton invariant mass spectrum.

At leading order in the $\lqcd/m_b$ expansion the $B$ meson decay rate is equal
to the $b$ quark decay rate.  Nonperturbative effects are suppressed by at
least two powers of $\lqcd / m_b$.  Corrections of order $\lqcd^2 / m_b^2$ are
characterized by two heavy quark effective theory (HQET) matrix
elements~\cite{incl,MaWi,Blok}, which are defined by
\begin{eqnarray}
\lambda_1 &=& \langle B(v)|\, \bar h_v^{(b)}\, (iD)^2\, h_v^{(b)}\,
  |B(v)\rangle/2m_B \,, \nonumber\\*
\lambda_2 &=& \langle B(v)|\, {g_s\over 2}\, \bar h_v^{(b)}\,
  \sigma_{\mu\nu} G^{\mu\nu}\, h_v^{(b)}\, | B(v) \rangle/6m_B \,.
\end{eqnarray}
These matrix elements also occur in the expansion of the $B$ and $B^*$
masses in powers of $\lqcd/m_b$,
\begin{equation}\label{massrelation}
m_B = m_b + \bar\Lambda 
  - {\lambda_1 + 3 \lambda_2 \over 2m_b} + \ldots \,, \qquad
m_{B^*} = m_b + \bar\Lambda 
  - {\lambda_1 - \lambda_2 \over 2m_b} + \ldots \,.
\end{equation}
Similar formulae hold for the $D$ and $D^*$ masses.  The parameters
$\bar\Lambda$ and $\lambda_1$ are independent of the heavy $b$ quark mass,
while there is a weak logarithmic scale dependence in $\lambda_2$.  The
measured $B^*-B$ mass splitting fixes $\lambda_2(m_b) = 0.12\,{\rm GeV}^2$.  At
${\cal O}(\lqcd^3/m_b^3)$ seven additional parameters arise in the
OPE~\cite{BDS,GK,m3corr}, and varying these parameters is often used to
estimate the theoretical uncertainty in the OPE~\cite{GK,m3corr,uncert}.

In inclusive semileptonic $B$ decay, for a particular hadronic final state $X$,
the maximum lepton energy is $E_\ell^{\rm (max)} = (m_B^2-m_X^2)/2m_B$ (in the
$B$ rest frame), so to eliminate charm background one must impose a cut $E_\ell
> (m_B^2-m_D^2)/2m_B$. The maximum lepton energy in semileptonic $b$ quark
decay is $m_b/2$, which is less than the physical endpoint $m_B/2$.  Their
difference, $\bar\Lambda/2$, is comparable in size to the endpoint region
$\Delta E_\ell^{\rm(endpoint)} = m_D^2 / 2m_B \simeq 0.33\,$GeV.  The effects
which extend the lepton spectrum beyond its partonic endpoint appear as
singular terms in the prediction for ${\rm d}\Gamma / {\rm d}E_\ell$ involving
derivatives of delta functions, $\delta^{(n)} (1-2E_\ell/m_b)$.  The lepton
spectrum must be smeared over a region of energies $\Delta E_\ell$ near the
endpoint before theory can be compared with experiment.  If the smearing region
$\Delta E_\ell$ is much smaller than $\lqcd$, then higher dimension operators
in the OPE become successively more important and the OPE is not useful for
describing the lepton energy spectrum.  For $\Delta E_\ell \gg \lqcd$, higher
dimension operators become successively less important and a useful prediction
for the lepton spectrum can be made using the first few terms in the OPE.  When
$\Delta E_\ell \sim \lqcd$, there is an infinite series of terms in the OPE
which are all equally important.  Since $\Delta E_\ell^{\rm(endpoint)}$ is
about $\lqcd$, it seems unlikely that predictions based on a few low dimension
operators in the OPE can successfully determine the lepton spectrum in this
region.  

It was shown in \cite{shape} that the leading singularities in the OPE may be
resummed into a nonperturbative light-cone distribution function $f(k_+)$ for
the heavy quark.  To leading order in $1/m_b$, the effects of the distribution
function may be included by replacing $m_b$ by $m_b^* \equiv m_b + k_+$, and 
integrating over the light-cone momentum
\begin{equation}
{{\rm d}\Gamma\over {\rm d}E_\ell} = \int {\rm d}k_+\, f(k_+)
  \left. {{\rm d}\Gamma_{\rm p}\over {\rm d}E_\ell} \right|_{m_b\to m_b^*} ,
\end{equation}
where ${\rm d}\Gamma_{\rm p}/{\rm d}E_\ell$ is the parton-level spectrum. 
Analogous formulae hold for other differential distributions \cite{dFN,DU}.  For
purposes of illustration, we will use a simple model for the structure function
given by the one-parameter ansatz~\cite{MN}
\begin{equation}\label{sfn}
f(k_+) = {32\over \pi^2 \Lambda}\, (1-x)^2\, 
  \exp\left[-{4\over \pi}(1-x)^2\right] \Theta(1-x) \,, 
\qquad  x \equiv {k_+\over \Lambda} \,,
\end{equation}
taking the model parameter $\Lambda=0.48\,$GeV, corresponding to $m_b=4.8\,$GeV.

\begin{figure}[t]
\centerline{\epsfysize=5.4truecm \epsfbox{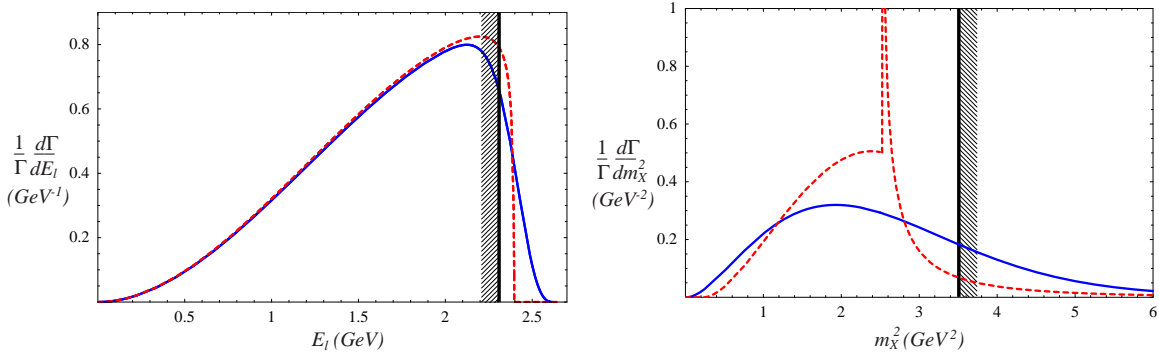} }
\caption[]{The lepton energy and hadronic invariant mass spectra.  The 
dashed curves are the $b$ quark decay results to ${\cal O}(\alpha_s)$, while 
the solid curves are obtained by smearing with the model distribution function 
$f(k_+)$ in Eq.~(\ref{sfn}).  The unshaded side of the vertical lines indicate 
the region free from charm background.  The area under each curve has been 
normalized to one.}
\label{twospectra}
\end{figure}

The charm background can also be eliminated by reconstructing the invariant
mass of the hadronic final state, $m_X$, since decays with $m_X < m_D$ must
arise from $b\to u$ transition.  While this analysis is challenging
experimentally, the $m_X < m_D$ cut allows a much larger fraction of $b\to u$
decays than the $E_\ell > (m_B^2-m_D^2)/2m_B$ constraint.  This is expected to
result in a reduction of the theoretical uncertainties~\cite{FLW,BDU}, although
both the lepton endpoint region, $E_\ell > (m_B^2-m_D^2)/2m_B$, and the low
hadronic invariant mass region, $m_X < m_D$, receive contributions from the
same set of hadronic final states (but with very different weights).  However,
the same nonperturbative effects which lead to the breakdown of predictive
power in the lepton endpoint region also give large uncertainties in the hadron
mass spectrum over the range $m_X^2 \sim \bar\Lambda m_b$~\cite{FLW}.  In
other words, nonperturbative effects yield formally ${\cal O}(1)$ uncertainties
in both cases, because numerically $m_D^2 \sim \lqcd m_B$.  The situation is
illustrated in Fig.~\ref{twospectra}.

The situation is very different for the lepton invariant mass spectrum. Decays
with $q^2 \equiv (p_\ell + p_{\bar\nu})^2 > (m_B - m_D)^2$ must arise from
$b\to u$ transition.  Such a cut forbids the hadronic final state from moving
fast in the $B$ rest frame, and so the light-cone expansion which gives rise to
the shape function is not relevant in this region of phase space.\footnote{The
fact that the $b$ quark distribution function is not relevant for large $q^2$
was pointed out in~\cite{BI} in the context of $\bar B\to X_s \ell^+\ell^-$
decay and in~\cite{DU} for semileptonic
$\bar B\rightarrow X_u$ decay.}  This is clear
from the kinematics: the difference between the partonic and hadronic values of maximum $q^2$
is
$m_B^2-m_b^2 \sim 2\bar\Lambda\, m_b$, and nonperturbative effects are only important in a
region of comparable size.  For example, the most singular term in the OPE at order
$(\lqcd/m_b)^3$ is of order $\left(\Lambda_{\rm QCD}/m_b\right)^3\, \delta(1-q^2/m_b^2)$. 
This contribution to the decay rate is not suppressed compared to the lowest order
term in the OPE only if the spectrum is integrated over a small region of width
$\Delta q^2 \sim \lqcd\, m_b$ near the endpoint.  This is the resonance region
where only hadronic final states with masses $m_X \sim \lqcd$ can contribute,
and the OPE is not expected to work anyway.  In contrast, nonperturbative
effects are important in the $E_\ell$ and $m_X^2$ spectra in a parametrically
much larger region where final states with masses $m_X^2 \sim \lqcd\, m_b$
contribute.\footnote{Similar arguments also show that the light-cone
distribution function is not relevant for small hadron energy, $E_X$, but it
does enter for $E_X$ near $m_b/2$.  If $m_b/2 - m_c \gg \lqcd$, then the
constraint $E_X<m_D$ in the $B$ rest frame would also give a model independent
determination of $|V_{ub}|$.}  The better behavior of the $q^2$ spectrum than
the $E_\ell$ and $m_X^2$ spectra is also reflected in the perturbation series. 
There are Sudakov double logarithms near the phase space boundaries in the
$E_\ell$ and $m_X^2$ spectra, whereas there are only single logarithms in the
$q^2$ spectrum.  

\begin{figure}[t]
\centerline{\epsfysize=5.4truecm \epsfbox{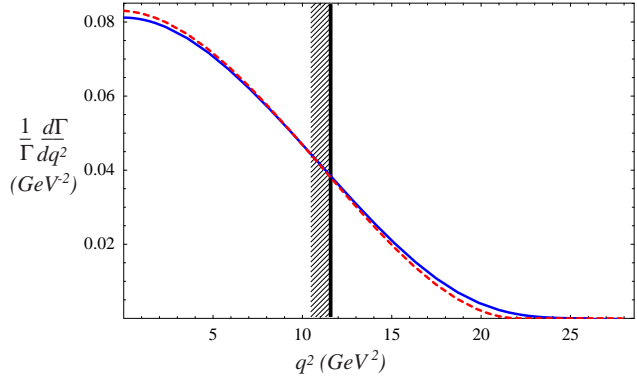} }
\caption[]{The lepton invariant mass spectrum to ${\cal O}(\alpha_s)$.  The
  meaning of the curves is the same as in Fig.~\ref{twospectra}.}
\label{qsqspectrum}
\end{figure}

The effect of smearing the $q^2$ spectrum with the model distribution function
in Eq.~(\ref{sfn}) is illustrated in Fig.~\ref{qsqspectrum}.  In accord with
our previous arguments, it is easily seen to be subleading over the region of
interest.  Table~I compares qualitatively the utility of the lepton energy, the
hadronic invariant mass, and the lepton invariant mass spectra for the
determination of~$|V_{ub}|$.  

\begin{table}[t]
\begin{tabular}{c|ccc}
Decay  &  Width of region without  &  Nonperturbative  
  &  ~~Fraction of $b\to u$~~ \\
~~~distribution~~~  &  charm background  &  region near endpoint  
  &  events included  \\ \hline
${\rm d}\Gamma/{\rm d}E_\ell$ &  $\Delta E_\ell =m_D^2/2 m_B$  
  &  $\Delta E_\ell \sim \lqcd$  &  $\sim 10\%$  \\
${\rm d}\Gamma/{\rm d}m_X^2$ &  $\Delta m_X^2 = m_D^2$  
  &  $\Delta m_X^2 \sim \lqcd\, m_b$  &  $\sim 80\% $  \\
${\rm d}\Gamma/{\rm d}q^2$ &  $\Delta q^2 =2 m_B m_D-m_D^2$  
  &  $\Delta q^2 \sim \lqcd\, m_b$  &  $\sim 20\%$ \\
\end{tabular} \vspace{4pt}
\caption[]{Comparison between the lepton energy, hadronic invariant mass, and
lepton invariant mass spectra for the determination of $|V_{ub}|$.  The region
dominated by nonperturbative effects is parametrically smaller than the region
without charm background only for the $q^2$ spectrum in the last row (viewing
$m_D^2\sim m_c^2 \sim \lqcd\, m_b$).  The last column gives rough numbers
corresponding to the plots in Figs.~1 and 2.} 
\end{table}

We now proceed to calculate the $\bar B\to X_u\ell\bar\nu$ decay rate with
lepton invariant mass above a given cutoff, working to a fixed order in the OPE
(i.e., ignoring the light-cone distribution function which is irrelevant for
our analysis).  The lepton invariant mass spectrum including the leading
perturbative and nonperturbative corrections is given by
\begin{eqnarray}\label{q2spec}
{1\over \Gamma_0}\, {{\rm d}\Gamma \over {\rm d}\qsh} &=&
  \bigg( 1 + {\lambda_1\over 2m_b^2} \bigg)\, 2\, (1-\qsh)^2\, (1+2\qsh) 
  + {\lambda_2\over m_b^2}\, (3 - 45\hat q^4 + 30\hat q^6) \nonumber\\*
&&{} + {\alpha_s(m_b) \over \pi}\, X(\qsh)
  + \bigg( {\alpha_s(m_b) \over \pi} \bigg)^2\, \beta_0\, Y(\qsh) + \ldots \,,
\end{eqnarray}
where $\qsh = q^2/m_b^2$, $\beta_0 = 11 - 2n_f/3$, and 
\begin{equation}
\Gamma_0 = {G_F^2\, |V_{ub}|^2\, m_b^5 \over 192\, \pi^3} 
\end{equation}
is the tree level $b\to u$ decay rate.  The ellipses in Eq.~(\ref{q2spec})
denote terms of order $(\lqcd/m_b)^3$ and order $\alpha_s^2$ terms not enhanced
by $\beta_0$.  The function $X(\qsh)$ is given analytically in Ref.~\cite{JK},
whereas $Y(\qsh)$ was computed numerically in Ref.~\cite{LSW}.  The order
$1/m_b^3$ nonperturbative corrections were computed in Ref.~\cite{m3corr}.  The
matrix element of the kinetic energy operator, $\lambda_1$, only enters the
$\hat q^2$ spectrum in a very simple form, because the unit operator and the
kinetic energy operator are related by reparameterization invariance~\cite{LM}.
Any quantity which can be written independent of the heavy quark velocity $v$
must depend only on the combination $(1 + \lambda_1 / 2m_b^2)$.  The $\qsh$
spectrum (and the total rate written in terms of $m_b$) are invariant under a
redefinition of $v$, but, for example, the lepton energy spectrum is not since
$E_\ell = v \cdot p_e$.  (Equivalently, the $\lambda_1$ term is a time-dilation
effect, and hence is universal in any quantity that is independent of the rest
frame of the $B$ meson \cite{incl}.)

We shall compute the fraction of $\bar B\to X_u \ell\bar\nu$ events with $q^2 >
\qcut^2$, $F(\qcut^2)$, as the relation between the total $\bar B\to X_u
\ell\bar\nu$ decay rate and $|V_{ub}|$ has been extensively discussed in the
literature~\cite{upsexp,burels}, and is known including the full $\alpha_s^2$
corrections~\cite{TvR}.  After integrating the spectrum in Eq.~(\ref{q2spec}),
we can eliminate the dependence on the $b$ quark mass in favor of the spin
averaged meson mass $\mbar = (m_B + 3 m_{B^*}) / 4 \simeq 5.313\,$GeV,
following~\cite{FLS}.  We find
\begin{eqnarray}\label{Fmeson}
F(\qcut^2) = 1 &-& 2 \Qcut^2 + 2 \Qcut^6 - \Qcut^8 
  - {4\bar\Lambda \over \mbar}\, \Big( \Qcut^2 - 3 \Qcut^6 + 2 \Qcut^8 \Big)
  - {6\bar\Lambda^2 \over \mbar^2}\,
  \Big( \Qcut^2 - 7 \Qcut^6 + 6 \Qcut^8 \Big) \nonumber\\
&+& {2\lambda_1 \over \mbar^2}\, \Big( \Qcut^2 - 3 \Qcut^6 + 2 \Qcut^8 \Big)
  - {12\lambda_2 \over \mbar^2}\,
  \Big( \Qcut^2 - 2 \Qcut^6 + \Qcut^8 \Big) \nonumber\\
&+& {\alpha_s(m_b) \over \pi}\, \tilde X(\Qcut^2)
  + \bigg( {\alpha_s(m_b) \over \pi} \bigg)^2\, \beta_0\, \tilde Y(\Qcut^2) 
  + \ldots \,,
\end{eqnarray}
where $\Qcut \equiv \qcut / \mbar$.  The functions $\tilde X(\Qcut^2)$ and
$\tilde Y(\Qcut^2)$ can be calculated from $X(\qsh)$ and $Y(\qsh)$.  Converting
to the physical $B$ meson mass has introduced a strong dependence on the
parameter $\bar{\Lambda}$, the mass of the light degrees of freedom in the $B$
meson.  
For $\qcut^2 = (m_B-m_D)^2 \simeq 11.6\,{\rm GeV}^2$, we find $F(11.6\, {\rm
GeV}^2) = 0.287 + 0.027 \alpha_s(m_b) - 0.016 \alpha_s^2(m_b) \beta_0 -
0.20\bar\Lambda/(1\ {\rm GeV}) - 0.02\bar\Lambda^2/(1\ {\rm GeV}^2) +
0.02\lambda_1/(1\ {\rm GeV}^2) - 0.13\lambda_2/(1\ {\rm GeV}^2) + \ldots$ . The order
$\alpha_s
\bar\Lambda$ term is negligible and has been omitted.  Using $\bar\Lambda = 0.4\,$GeV,
$\lambda_1 = -0.2\,{\rm GeV}^2$~\cite{gremmetal} and $\alpha_s(m_b) = 0.22$, we obtain
$F(11.6\, {\rm GeV}^2) = 0.186$.  There are several sources of uncertainties in the value for
$F$. The perturbative uncertainties are negligible, as can be seen from the
size of the ${\cal O} (\alpha_s^2 \beta_0)$ contributions. At the present time
there is a sizable uncertainty since $\bar\Lambda$ is not known accurately. In
the future, a $\pm50\,$MeV error in $\bar\Lambda$ will result in a $\pm5\%$
uncertainty in $F$.  Finally, uncertainties from $1/m_b^3$ operators can be
estimated by varying the matrix elements of the dimension six operators within
the range expected by dimensional analysis, as discussed in detail
in~\cite{GK,m3corr,uncert}.  This results in an additional $\pm 4\%$
uncertainty in $F$.  We note that this is a somewhat ad hoc procedure, since
there is no real way to quantify the theoretical error due to unknown higher
order terms.  Therefore, these estimates should be treated as nothing more than
(hopefully) educated guesses.  They do allow, however, for a consistent
comparison of the uncertainties in different quantities.  

If $\qcut^2$ has to be chosen larger, then the uncertainties increase.  For
example, for $\qcut^2 = 15\,{\rm GeV}^2$, we obtain $F(15\, {\rm GeV}^2) =
0.158 + 0.024 \alpha_s(m_b)-0.012 \alpha_s^2(m_b)\beta_0 - 0.18\bar\Lambda/(1\ {\rm GeV}) +
0.01\bar\Lambda^2/(1\ {\rm GeV}^2) + 0.02\lambda_1/(1\ {\rm GeV}^2) - 0.13\lambda_2/
(1\ {\rm GeV}^2) + \ldots
\simeq 0.067$, using the previous values of $\bar\Lambda$ and $\lambda_1$.  The perturbative
uncertainties are still negligible, while the uncertainty due to a $\pm50\,$MeV
error in $\bar\Lambda$ and unknown dimension six matrix elements increase to
$\pm14\%$ and $\pm 13\%$, respectively.  (This uncertainty may be reduced using
data on the rare decay $\bar B \to X_s \ell^+\ell^-$ in the large $q^2$ region,
as discussed below.)

Another possible method to compute $F(\qcut^2)$ uses the upsilon
expansion~\cite{upsexp}.  By expressing $\hat q^2$ in terms of the $\Upsilon$
mass instead of $\mbar$, the dependence of $F(\qcut^2)$ on $\bar\Lambda$ and
$\lambda_1$ is eliminated.  Instead, the result is sensitive to unknown
nonperturbative contributions to $m_\Upsilon$.  The uncertainty related to
these effects can be systematically taken into account and has been estimated
to be small~\cite{upsexp}.  One finds,
\begin{equation}\label{Vubups}
|V_{ub}| = (3.04 \pm 0.06 \pm 0.08) \times 10^{-3}\,
  \left( {{\cal B}(\bar B\to X_u \ell\bar\nu)|_{q^2 > \qcut^2} \over 
  0.001 \times F(\qcut^2) }\, {1.6\,{\rm ps}\over\tau_B} \right)^{1/2} \,.
\end{equation}
The errors explicitly shown in Eq.~(\ref{Vubups}) are the estimates of the
perturbative and nonperturbative uncertainties in the upsilon expansion,
respectively.

\begin{figure}[t]
\centerline{\epsfysize=5.4truecm \epsfbox{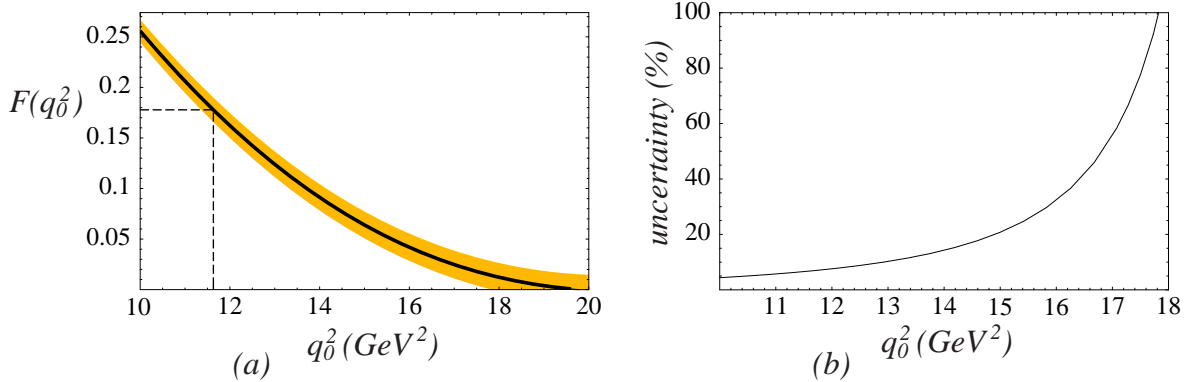} }
\caption[]{(a) The fraction of $\bar B\to X_u \ell\bar\nu$ events with $q^2 >
\qcut^2$, $F(\qcut^2)$, to order $\epsilon_{\rm BLM}^2$ and $\lqcd^2/m_b^2$ in
the upsilon expansion.  The shaded region is the uncertainty due to $\lqcd^3 /
m_b^3$ terms, as discussed in the text.  The dashed line indicates the lower
cut $\qcut^2 = (m_B-m_D)^2 \simeq 11.6\,{\rm GeV}^2$, which corresponds to $F =
0.178 \pm 0.012$. (b) The estimated uncertainty in $F(\qcut^2)$ due to $\lqcd^3
/ m_b^3$ terms as a percentage of $F(\qcut^2)$. }
\label{fractionplot}
\end{figure}

For $\qcut^2 = (m_B-m_D)^2$ we find $F(11.6\, {\rm GeV}^2) = 0.168 + 0.016
\epsilon + 0.014 \epsilon_{\rm BLM}^2 - 0.17 \lambda_2 + \ldots \simeq 0.178$,
where $\epsilon \equiv 1$ shows the order in the upsilon expansion.  This
result is in good agreement with 0.186 obtained from Eq.~(\ref{Fmeson}). The
uncertainty due to $\bar\Lambda$ is absent in the upsilon expansion, however
the size of the perturbative corrections has increased.  The uncertainties due
to $1/m_b^3$ operators is estimated to be $\pm 7\%$.  For $\qcut^2 = 15\,{\rm
GeV}^2$, we obtain $F(15\,{\rm GeV}^2) = 0.060 + 0.011 \epsilon + 0.011
\epsilon_{\rm BLM}^2 - 0.14 \lambda_2 + \ldots \simeq 0.064$, which is again in
good agreement with 0.067 obtained earlier.  For this value of $\qcut^2$, the
$1/m_b^3$ uncertainties increase to $\pm 21\%$.  $F(\qcut^2)$ calculated in the
upsilon expansion is plotted in Fig.~\ref{fractionplot}, where the shaded
region shows our estimate of the uncertainty due to the $1/m_b^3$ corrections.

Concerning experimental considerations, measuring the $q^2$ spectrum requires
reconstruction of the neutrino four-momentum, just like measuring the hadronic
invariant mass spectrum.  Imposing a lepton energy cut, which may be required
for this technique, is not a problem.  The constraint $q^2 > (m_B-m_D)^2$
automatically implies $E_\ell > (m_B-m_D)^2/2m_B \sim 1.1\,$GeV in the $B$ rest
frame.  Even if the $E_\ell$ cut has to be slightly larger than this, the
utility of our method will not be affected, but a dedicated calculation
including the affects of arbitrary $E_\ell$ and $q^2$ cuts may be warranted.

If experimental resolution on the reconstruction of the neutrino momentum
necessitates a significantly larger cut than $\qcut^2 = (m_B-m_D)^2$, then the
uncertainties in the OPE calculation of $F(\qcut^2)$ increase.  In this case,
it may instead be possible to obtain useful model independent information on
the $q^2$ spectrum in the region $q^2 > m_{\psi(2S)}^2 \simeq 13.6\,{\rm
GeV}^2$ from the $q^2$ spectrum in the rare decay $\bar B \to X_s
\ell^+\ell^-$, which may be measured in the upcoming Tevatron Run-II.  There
are four contributions to this decay rate, proportional to the combination of
Wilson coefficients $\widetilde C_9^2$, $C_{10}^2$, $C_7 \widetilde C_9$, and
$C_7^2$.  $\widetilde C_9$ is a $q^2$-dependent effective coefficient which
takes into account the contribution of the four-quark operators.  Its
$\qsh$-dependence yields negligible uncertainties if we use a mean $\widetilde
C_9$ obtained by averaging it in the region $0.5 < \qsh < 1$ weighted with the
$b$ quark decay rate $(1-\qsh)^2\, (1+2\qsh)$.  The resulting numerical values
of the Wilson coefficients are $\widetilde C_9 = 4.47 + 0.44\,i$, $C_{10} =
-4.62$, and $C_7 = -0.31$, corresponding to the scale $\mu = m_b$.  In the $q^2
> m_{\psi(2S)}^2$ region the $C_7^2$ contribution is negligible, and the $C_7
\widetilde C_9$ term makes about a 20\% contribution to the rate.  For the
$\widetilde C_9^2 + C_{10}^2$ contributions nonperturbative effects are
identical to those which occur in $\bar B\to X_u \ell\bar\nu$ decay, up to
corrections suppressed by $|\widetilde C_9 + C_{10}| / |\widetilde C_9 -
C_{10}| \sim 0.02$.  Therefore, the relation
\begin{equation}\label{raredecay}      
{ {\rm d}\Gamma(\bar B \to X_u\ell\bar\nu) / {\rm d}\qsh \over 
  {\rm d}\Gamma(\bar B \to X_s\ell^+\ell^-) / {\rm d}\qsh }
= {|V_{ub}|^2 \over |V_{ts} V_{tb}|^2}\, {8\pi^2 \over \alpha^2}\,
  { 1 \over |\widetilde C_9|^2 + |C_{10}|^2 
  + 12\, {\rm Re}\, (C_7 \widetilde C_9) / (1+2\qsh) } \,,
\end{equation}
is expected to hold to a very good accuracy.  There are several sources of
corrections to this formula which need to be estimated: i)~nonperturbative
effects that enter the $C_7 \widetilde C_9$ term differently, ii)~mass effects
from the strange quark and muon, iii)~higher $c \bar c$ resonance contributions
in $\bar B\to X_s \ell^+\ell^-$, and iv)~scale dependence.  Of these, i) and
ii) are expected to be small unless $q^2$ is very close to $m_B^2$.  The
effects of iii) have also been estimated to be at the few percent
level~\cite{BI}, although these uncertainties are very hard to quantify and
could be comparable to the $\pm8\%$ scale dependence~\cite{BuMu} of the $\bar
B\to X_s \ell^+\ell^-$ rate.  Integrating over a large enough range of $q^2$,
$\qcut^2 < q^2 < m_B^2$ with $m_{\psi(2S)}^2 < \qcut^2 \lesssim 17\,{\rm
GeV}^2$, the result implied by Eq.~(\ref{raredecay}),
\begin{equation}
{ {\cal B}(\bar B \to X_u\ell\bar\nu)|_{q^2 > \qcut^2} \over
  {\cal B}(\bar B \to X_s\ell^+\ell^-)|_{q^2 > \qcut^2} }
= {|V_{ub}|^2 \over |V_{ts} V_{tb}|^2}\, {8\pi^2 \over \alpha^2}\,
  { 1 \over |\widetilde C_9|^2 + |C_{10}|^2 
  + 12\, {\rm Re}\, (C_7 \widetilde C_9)\, B(\qcut^2) } \,,
\end{equation}
is expected to hold at the $\sim15\%$ level.  Here $\Qcut \equiv \qcut /
\mbar$, and $B(\qcut^2) = 2 / [3 (1+\Qcut^2)] - 4 (\bar\Lambda/\mbar)\, \Qcut^2
/ [3 (1+\Qcut^2)^2] + \ldots$ .  For $\qcut^2$ significantly above
$(m_B-m_D)^2$, this formula may lead to a determination of $|V_{ub}|$ with
smaller theoretical uncertainty than the one using the OPE calculation of
$F(\qcut^2)$.

In conclusion, we have shown that the $q^2$ spectrum in inclusive semileptonic
$\bar B \to X_u \ell \bar\nu$ decay gives a model independent determination of
$|V_{ub}|$ with small theoretical uncertainty.  Nonperturbative effects are
only important in the resonance region, and play a parametrically suppressed
role when ${\rm d}\Gamma/{\rm d}q^2$ is integrated over $q^2>(m_B-m_D)^2$,
which is required to eliminate the charm background.  This is a qualitatively
better situation than the extraction of $|V_{ub}|$ from the endpoint region of
the lepton energy spectrum, or from the hadronic invariant mass spectrum.

\acknowledgements

We thank Craig Burrell for discussions and Adam Falk for comments on the
manuscript.
This work was supported in part by the Natural Sciences and Engineering
Research Council of Canada and the Sloan Foundation.  Fermilab is operated by
Universities Research Association, Inc., under DOE contract DE-AC02-76CH03000.

\end{document}